\documentclass[conference]{IEEEtran}
\IEEEoverridecommandlockouts

\usepackage{cite}
\usepackage{amsmath,amssymb,amsfonts}
\usepackage{algorithmic}
\usepackage{graphicx}
\usepackage{textcomp}
\usepackage{xcolor}
\usepackage{svg}
\usepackage[hidelinks]{hyperref}
\def\BibTeX{{\rm B\kern-.05em{\sc i\kern-.025em b}\kern-.08em
    T\kern-.1667em\lower.7ex\hbox{E}\kern-.125emX}}

\usepackage{subcaption}
\usepackage{url}
\usepackage{booktabs}

\usepackage[acronym, shortcuts]{glossaries-extra}
\glssetcategoryattribute{acronym}{nohyper}{true}
\setabbreviationstyle[acronym]{long-short}

\newacronym{ai}{AI}{Artificial Intelligence}
\newacronym{llm}{LLM}{Large Language Models}
\newacronym{cvss}{CVSS}{Common Vulnerability Scoring System}
\newacronym{cve}{CVE}{Common Vulnerabilities and Exposure}
\newacronym{cti}{CTI}{Cyber-Threat Intelligence}
\newacronym{rag}{RAG}{Retrieval-Augmented Generation}
\newacronym{cwe}{CWE}{Common Weakness Enumeration}
\newacronym{rf}{RF}{Random Forest}
\newacronym{ml}{ML}{Machine Learning}
\newacronym{av}{AV}{Attack Vector}
\newacronym{ac}{AC}{Attack Complexity}
\newacronym{pr}{PR}{Privileges Required}
\newacronym{ui}{UI}{User Interaction}
\newacronym{s}{S}{Scope}
\newacronym{c}{C}{Confidentiality}
\newacronym{i}{I}{Integrity}
\newacronym{a}{A}{Availability}

\begin{document}

\title{Can LLMs Classify CVEs? Investigating LLMs Capabilities in Computing CVSS Vectors}

\author{\IEEEauthorblockN{Francesco Marchiori}
\IEEEauthorblockA{\textit{Department of Mathematics} \\
\textit{University of Padova}\\
Padua, Italy \\
francesco.marchiori.4@phd.unipd.it}
\and
\IEEEauthorblockN{Denis Donadel}
\IEEEauthorblockA{\textit{Department of Computer Science} \\
\textit{University of Verona}\\
Verona, Italy \\
denis.donadel@univr.it}
\and
\IEEEauthorblockN{Mauro Conti}
\IEEEauthorblockA{\textit{Department of Mathematics} \\
\textit{University of Padova}\\
Padua, Italy \\
mauro.conti@unipd.it}
}

\maketitle

\begin{abstract}
Common Vulnerability and Exposure (CVE) records are fundamental to cybersecurity, offering unique identifiers for publicly known software and system vulnerabilities. Each CVE is typically assigned a Common Vulnerability Scoring System (CVSS) score to support risk prioritization and remediation. However, score inconsistencies often arise due to subjective interpretations of certain metrics. As the number of new CVEs continues to grow rapidly, automation is increasingly necessary to ensure timely and consistent scoring. While prior studies have explored automated methods, the application of Large Language Models (LLMs), despite their recent popularity, remains relatively underexplored.

In this work, we evaluate the effectiveness of LLMs in generating CVSS scores for newly reported vulnerabilities. We investigate various prompt engineering strategies to enhance their accuracy and compare LLM-generated scores against those from embedding-based models, which use vector representations classified via supervised learning. Our results show that while LLMs demonstrate potential in automating CVSS evaluation, embedding-based methods outperform them in scoring more subjective components, particularly confidentiality, integrity, and availability impacts. These findings underscore the complexity of CVSS scoring and suggest that combining LLMs with embedding-based methods could yield more reliable results across all scoring components.
\end{abstract}


\section{Introduction}\label{sec:intro}

In cybersecurity, managing and addressing vulnerabilities is crucial to maintaining systems and data integrity.
The \ac{cve} system is a standardized method for identifying and cataloging publicly known security vulnerabilities, providing a common reference point for organizations to communicate risks.
Alongside \ac{cve}, the \ac{cvss} plays a key role in evaluating the severity of these vulnerabilities, enabling organizations to prioritize remediation efforts.
It is generated in the form of a vector of elements representing different features of the \ac{cve} that can be summarized in a single score (\ac{cvss} score), making it a valuable index for prioritization.  
Together, \ac{cve} and \ac{cvss} offer essential tools for identifying, understanding, and addressing vulnerabilities consistently and structured, helping organizations mitigate potential security threats more effectively.

Generating the \ac{cvss} vector (and consequently \ac{cvss} scores) from \acp{cve} is a demanding task for analysts~\cite{wunder2024shedding}.
Not only the number of \acp{cve} has increased by 38\% from 2023 to 2024~\cite{Kaaviya2025}, but the final score can also differ based on the subjective nature of some of the \ac{cvss} vector, such as the level of confidentiality, integrity, and availability risk.
With more than 135 new \acp{cve} published each day~\cite{Fortress2024}, automation is becoming increasingly essential to streamline the process of scoring vulnerabilities, ensuring faster response times, reducing human error, and enhancing the overall efficiency of cybersecurity operations.

\ac{ai} may help in this task by providing suggestions to users or ultimately generating the whole vector.
Some experiments using \acp{llm} showed promising results in specific contexts~\cite{ghosh2025cve}, but the applicability of general-purpose \acp{llm} for the task has never been investigated.

\textit{Contributions.}
In this paper, we investigate the applicability of \acp{llm} in generating \ac{cvss} vectors from \ac{cve} descriptions to facilitate and assist analysis.
We conduct several experiments in different conditions and with various models.
We also explore the feasibility of using text embedding models and traditional \ac{ml} classification models to generate the same vectors, providing a training-based approach.
The main contributions of this paper can be summarized as follows.
\begin{itemize}
    \item We provide the first comprehensive analysis on the capabilities of \acp{llm} in generating \ac{cvss} vectors and scores for \acp{cve}.
    \item We show how Gemma3 is the best-performing model for the task, reaching an accuracy of up to 0.98 in the best-performing CVSS vector elements.
    \item We discussed the pros and cons of \ac{llm} versus an embedding approach. We show how a hybrid approach can strengthen the automatic CVSS generation reaching a mean accuracy of 0.84 through all the CVSS elements. 
    \item We make our code and implementation open-source: \texttt{\url{https://github.com/spritz-group/LLM-CVSS}}.
\end{itemize}
\section{Related Works}\label{sec:related}

Since the advent of the first \acp{llm}, researchers have started employing them in a wide range of fields. Different works propose \ac{llm}-based solutions for malware detection, program repair, anomaly detection, fuzzing, vulnerability detection, and other cybersecurity-related tasks~\cite{zhang2025llms}. Another sector that founds applicability of \acp{llm} is the \ac{cti}, where generative models can be employed to enhance different processes.
LOCALINTEL is a framework that enables \ac{cti} organization-specific contextualization~\cite{mitra2024localintel}. It enables the generation of contextualized information from a repository of global threats (e.g., \acp{cve}) together with a local knowledge graph containing, for instance, configurations of the employed devices. Similarly, Perrina et al.~\cite{perrina2023agir} developed AGIR, a tool to automatically generate \ac{cti} reports. 

Language models have been used to extract intelligence from darknet forums or other unstructured data sources~\cite{siracusano2023time}.
AttacKG+, a fully automated framework proposed by Zhang et al., allows for the construction of knowledge graphs from \ac{cti} using \acp{llm}~\cite{zhang2024attackg}.  
Fayyazi et al. proposed a solution to summarize Tactics, Techniques, and Procedures (TTPs) from the MITRE ATT\&CK framework boosting \ac{llm} with \ac{rag}~\cite{fayyazi2024advancing}.

Some works proposed also the employment of LLMs as security and network experts~\cite{ullah2024llms}. Lin et al. designed a solution that allows the retrieval of \acp{cve} and \ac{cwe} from a vulnerability description, and use it for vulnerabilities mitigation~\cite{lin2023hw}. Donadel et al.~\cite{donadel2024can} investigated the capabilities of \acp{llm} in understanding and answering questions related to a provided network architecture. However, none of these tools is specifically designed or tested on \acp{cve} classification and ranking.

In cybersecurity, embedding models have proven effective for automating the classification and prioritization of vulnerabilities.
By converting vulnerability descriptions into vector representations, these models support tasks such as \ac{cvss} scoring and \ac{cwe} mapping.
CVSS-BERT, for example, predicts \ac{cvss} metrics using BERT-based classifiers with explainable outputs~\cite{shahid2021cvss}, while V2W-BERT links \acp{cve} to \acp{cwe} through transformer-based learning~\cite{das2021v2w}.
Such methods highlight the potential of embeddings to streamline vulnerability assessment and support informed security decisions.
\section{Background}\label{sec:background}

Common Vulnerabilities and Exposures (CVEs) are publicly disclosed cybersecurity vulnerabilities that have been cataloged to provide a standardized reference for identifying and discussing specific issues.
Each CVE is assigned a unique identifier and is maintained by the MITRE Corporation\footnote{\url{https://cve.mitre.org/}} as part of a broader effort to facilitate information sharing across the cybersecurity community.

The Common Vulnerability Scoring System (CVSS) is a standardized framework for rating the severity of security vulnerabilities. It is widely used by organizations and researchers to assess and prioritize the risks associated with specific CVEs. CVSS provides a numerical score that reflects the potential impact and exploitability of a vulnerability.

CVSS scores are computed directly from the CVSS vector, which encodes a set of base metrics describing key characteristics of the vulnerability. These metrics include the \ac{av}, \ac{ac}, \ac{pr}, \ac{ui}, \ac{s}, and the impacts on \ac{c}, \ac{i}, and \ac{a}. An example of a complete vector is the following:
\begin{equation*}
\text{\small\texttt{CVSS:3.1/AV:N/AC:H/PR:L/UI:R/S:C/C:L/I:L/A:H}}
\end{equation*}
Each metric is assigned a specific value, and these values are combined using a defined formula to compute the Base Score, which ranges from 0.0 to 10.0. The formula consists of two main subscores: the Exploitability Subscore, derived from AV, AC, PR, and UI; and the Impact Subscore, calculated from C, I, A, and adjusted based on Scope. 
These subscores are combined using a deterministic set of equations defined in the CVSS specification (currently CVSS v3.1), ensuring that the final score reflects both the ease of exploitation and the potential impact of the vulnerability~\cite{cvss-specifications}.

\section{Methodology}\label{sec:methodology}

This paper aims to provide an overall evaluation of the capabilities of general-purpose \acp{llm} in classifying \acp{cve}. In particular, we aim to provide an answer to the following research questions.
\begin{enumerate}
    \renewcommand{\labelenumi}{\textbf{RQ\theenumi}}
    \item \label{rq1} Are general-purpose \acp{llm} capable of classifying \ac{cve} description generating the related \ac{cvss} vector?
    \item \label{rq2} Are there any prompt engineering tricks to enhance runtime performance?
    \item \label{rq3} How ``traditional'' classification techniques based on textual embeddings perform with respect to \acp{llm}?
\end{enumerate}

We set two scenarios to answer our research queries: \textit{vanilla} and \textit{embeddings}. The former setup directly employs general-purpose \ac{llm} without any fine-tuning, while the latter employs different extractors to generate embeddings from \ac{cve} descriptions and traditional \ac{ml} models to classify them.

\begin{figure}[htbp]
    \centering
    \begin{subfigure}{0.95\columnwidth}
        \includegraphics[width=\linewidth]{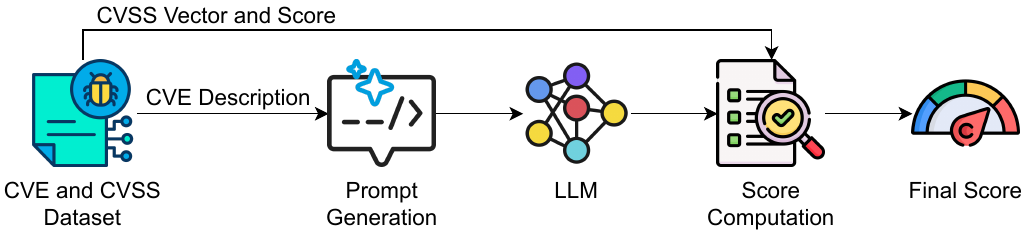}
        \caption{\textit{Vanilla} architecture.}
        \label{subfig:vanilla-arch}
    \end{subfigure}
    \begin{subfigure}{0.95\columnwidth}
        \includegraphics[width=\linewidth]{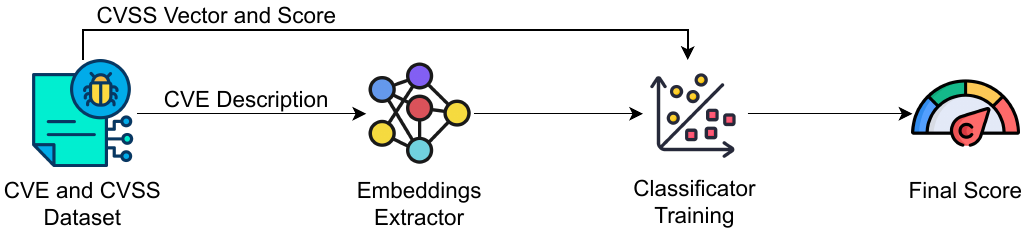}
        \caption{\textit{Embeddings} architecture.}
        \label{subfig:embeddings-arch}
    \end{subfigure}
    \caption{Architecture employed in our experiments.}
    \label{fig:architectures}
\end{figure}

\subsection{Vanilla \acp{llm}}
\label{subsec:vanilla}

In the base scenario, we consider only plain and untouched LLMs. 
A prompt is generated by taking a CVE description as input, which is then fed to the LLM under test, asking to generate the CVSS vector, as shown in Figure~\ref{subfig:vanilla-arch}.
Then, the output is fed to a simple score calculator, which will compute the success rate based on the ground-truth \ac{cvss} information. 

We performed various experiments in this scenario while changing and improving the employed prompt.
The base prompt asked the model to return the complete \ac{cvss} vector for one \ac{cve} at the time. 
Then, we tried a few-shot approach to explain the task to the model better.
Our prompt is enhanced for this experiment with a list of 3 examples preceding the CVE description under investigation.
We selected these examples in such a way that they maximize the variability between the CVSS components while keeping the most commonly known CVEs to ensure sufficient representation in the original training dataset.
We also limit to only three examples given the possibly limited capabilities of small \acp{llm} under test, avoiding the risks that can occur with a smaller context window.
Moreover, since the CVE description is often short and misses some vulnerability details, we test the effect of enhancing the description by adding \ac{cwe} descriptions.
The hypothesis is that, by knowing the specific \ac{cwe} exploited by a \ac{cve}, it would be possible to evaluate better the risk on specific \ac{cvss} vectors.
Thus, we extract the \ac{cwe} identifier from each \ac{cve} entry and enrich it with additional information retrieved from the official MITRE database\footnote{Example: \url{https://cwe.mitre.org/data/definitions/89.html}.}.
Specifically, we include the \ac{cwe} description, common consequences, and potential mitigations.
These fields were chosen because they are consistently available across most \ac{cwe} entries and provide valuable insights into the nature and impact of the vulnerability.
We also experiment with a prompt asking the \ac{llm} to generate a single vector component. Therefore, eight queries are required to compute the whole \ac{cvss} vector in this case. 

Finally, we prompt the \acp{llm} with the \ac{cve} description and ask them to directly generate the overall \ac{cvss} score without classifying each vector component.
This approach allows us to assess the models' ability to estimate the risk associated with a given \ac{cve}, providing insight into their potential usefulness for vulnerability prioritization.

Regardless of the specific task, each input prompt is constructed by providing information on each \ac{cvss} component (in the case of single-component classification, only the relevant one is included).
We also include an example of the output format that the \ac{llm} should use, and we highlight the need to generate only the relevant information (i.e., vector or component) in the specified format.
Since some of the \acp{llm} used also have reasoning capabilities, we filter the response by removing possible \texttt{<think>} tags in the response.
Furthermore, we consider cases where the model provides an answer in a different format, e.g., by including additional reasonings or specifying the label (e.g., ``Network'' instead of ``N'').
We do this by defining a hierarchy of Regex rules, which process and filter the response into the allowed labels for the specific component.
All prompts and rules used are shown in our repository.

\subsection{Embeddings}

Another completely different approach considers the usage of embeddings. Embeddings are numerical vector representations of text that capture their semantic meaning. They allow models to compare, relate, and reason about language by placing similar concepts close together in a high-dimensional space~\cite{kusner2015word}. 
As shown in Figure~\ref{subfig:embeddings-arch}, we employed different models to generate embeddings from \ac{cve} descriptions. 
Then, we trained an \ac{ml} classifier to classify each of the \ac{cvss} vector elements, allowing for a reconstruction of the overall vector. It can then be used to compute the \ac{cvss} score as per specification~\cite{cvss-specifications}.
To enhance the capabilities of the models, we also considered the addition of the same \ac{cwe} information detailed in Section~\ref{subsec:vanilla}.
Each of these parameters is then embedded and considered as a separate feature we can use for classification and analysis.

\subsection{Experimental Setup}

\begin{figure}[htbp]
    \centering
    \includegraphics[width=.95\columnwidth]{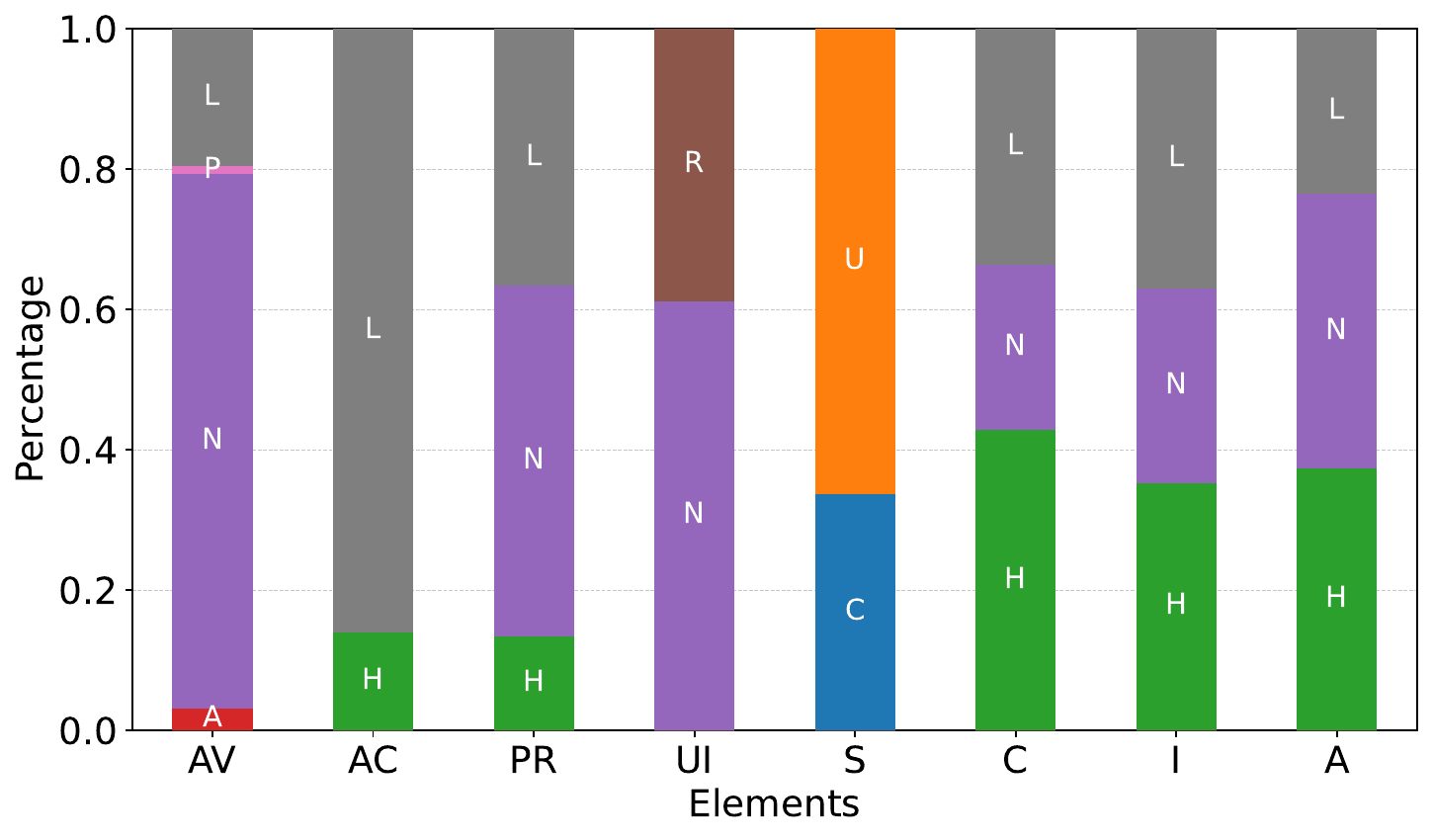}
    \caption{Distribution of \ac{cvss} vector elements in our dataset.\protect\footnotemark}
    \label{fig:dataset-summary}
\end{figure}

We collected \acp{cve} from the official \ac{cve} Github repository\footnote{\url{https://github.com/CVEProject/cvelistV5/}}.
We kept only entries with an associated \ac{cvss} v3.1 vector and score, which represents an updated and widely used version of the indicator.
Indeed, although CVSS v4.0 exists, not every entry in the dataset contains it.
After this preprocessing, we end up with a dataset of 45k entries of \acp{cve} from 2002 to 2025.
Figure~\ref{fig:dataset-summary} summarizes the data distribution inside the dataset.
In the embedding scenario, we utilized the entire dataset for training, while for the vanilla scenario, which does not involve any training phase, we restricted the test set to the most recent 1000 \acp{cve}. This approach ensures that the \acp{cve} in the test set were not included in the training data of the \acp{llm}, as all data entries date from March 2025 onward.

\footnotetext{Labels:  
AV (Attack Vector): Network (N), Adjacent (A), Local (L), Physical (P); 
AC (Attack Complexity): Low (L), High (H);  
PR (Privileges Required): None (N), Low (L), High (H);  
UI (User Interaction): None (N), Required (R);  
S (Scope): Unchanged (U), Changed (C);  
C/I/A (Confidentiality, Integrity, Availability): None (N), Low (L), High (H).}

In the vanilla scenario, we employed different \acp{llm}, both proprietary and open-weights.
In particular, we considered DeepSeek R1 (8B parameters)~\cite{guo2025deepseek}, Gemma 3 (12B parameters)~\cite{team2024gemma}, GPT 4o Mini~\cite{hurst2024gpt}, Llama 3.2 (3B parameters)~\cite{touvron2023llama}, Mistral~\cite{jiang2024mixtral}, Phi 4~\cite{abdin2024phi}, and Qwen 2.5 (14B parameters)~\cite{yang2024qwen2}.

In the embeddings scenarios, instead, we employed three different models to extract the embeddings: all-miniLM (L6, version 2), nomic, and ATTACK-BERT.\footnote{A version of the widely used BERT model fine-tuned on cybersecurity tasks: \url{https://huggingface.co/basel/ATTACK-BERT}}
For classification, we tested different models (XGBoost, Random Forest, Logistic Regression, and some simple Neural Networks) and ended up using XGBoost~\cite{chen2016xgboost}, which shows better results.  
We act a split of 80\%/20\% between training and testing data. 
We employed accuracy as the metric to show our results, representing the ratio between the number of correctly classified samples and the total number of samples for a specific vector element.
\section{Results}\label{sec:results}

In this section, we show the results of the experiments in both the vanilla and embedding case. Moreover, we propose a comparison of the best approaches in two scenarios. 

\subsection{Vanilla LLMs}
We start by analyzing the most basic scenario possible, i.e., asking an \ac{llm} to fill each of the \ac{cvss} vector elements together.
Results of the vanilla scenarios are shown in Table~\ref{tab:vanilla-results}. 
Generally, we can see how all the considered \acp{llm} performed well, despite some important differences between vector elements.
In particular, while the \ac{av} is detected correctly by almost all the considered models, \ac{pr} is more tricky.
Regarding the impact metrics (C, I, A), the results are not satisfying.
While \ac{c} and \ac{i} exhibit reasonable accuracy when using the Gemma 3 model, \ac{a} achieve only 0.47 in the best case. 
However, Gemma 3 is overall the best-performing model. 

\begin{table}[htbp]
\centering
\caption{Accuracies in the vanilla scenario for different \acp{llm} in generating the \ac{cvss} vector in one single shot.}
\label{tab:vanilla-results}
\resizebox{\columnwidth}{!}{%
\begin{tabular}{l|cccccccc}
\toprule
\textbf{Model} & \textbf{AV}   & \textbf{AC}   & \textbf{PR}   & \textbf{UI}   & \textbf{S}    & \textbf{C}    & \textbf{I}    & \textbf{A}    \\ \midrule
DeepSeek       & 0.91          & 0.86          & 0.46          & 0.63          & 0.55          & 0.35          & 0.29          & 0.33          \\
Gemma 3        & 0.95          & \textbf{0.92} & \textbf{0.48} & \textbf{0.95} & \textbf{0.82} & 0.62          & \textbf{0.70} & \textbf{0.47} \\
GPT 4o Mini & 0.97 &	\textbf{0.92} & 	\textbf{0.48} &	0.94 &	0.65 & 	0.35 & 	0.31 & 	0.34 \\
Llama 3.2      & 0.53          & 0.52          & 0.41          & 0.66          & 0.57          & 0.23          & 0.22          & 0.24          \\
Mistral        & \textbf{0.98} & \textbf{0.92} & 0.47          & 0.71          & 0.41          & 0.62          & 0.61          & 0.33          \\
Phi 4          & 0.95          & \textbf{0.92} & \textbf{0.48} & 0.76          & 0.45          & 0.24          & 0.23          & 0.29          \\
Qwen 2.5       & 0.97          & 0.91          & 0.47          & 0.76          & 0.61          & \textbf{0.63} & 0.22          & 0.26          \\
\bottomrule
\end{tabular}
}
\end{table}

\begin{figure}[htbp]
    \centering
    \includegraphics[width=.95\columnwidth]{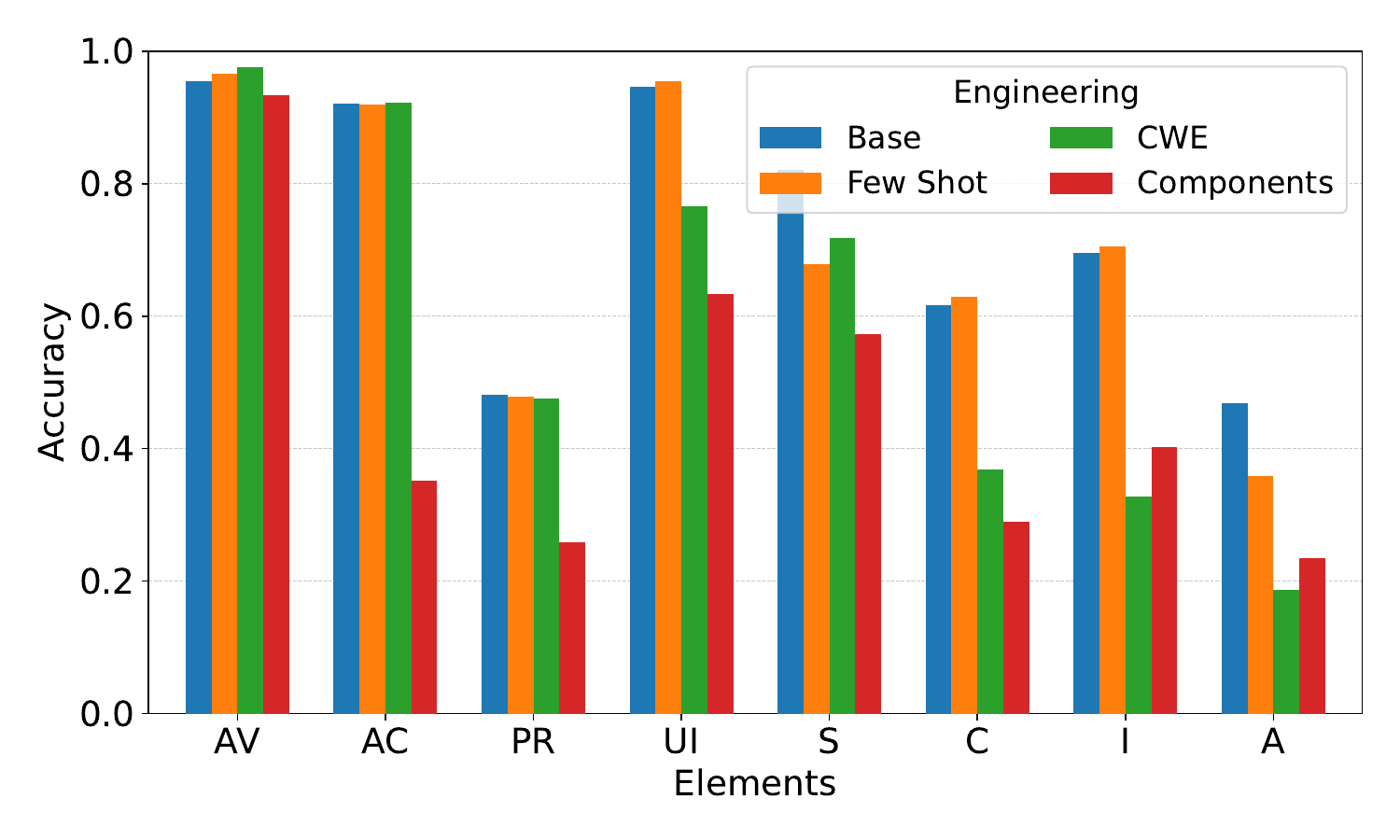}
    \caption{Accuracy with different prompt engineering techniques using the best performing model (Gemma 3).}
    \label{fig:prompt-eng}
\end{figure}

Moreover, we tested some prompt engineering enhancements to see if the \acp{llm} might benefit from more information in the prompt or for a more precise question. 
Results are shown in Figure~\ref{fig:prompt-eng}.
In the case of \ac{av}, all the prompts show similar results, with a negligible advancement of the prompt containing \acp{cwe}. 
Similarly, \ac{ac} and \ac{pr} show steady accuracies for \ac{cwe} and few shots compared to the base prompt, while asking one element at a time significantly decreases the accuracies.
In the others, instead, \ac{cwe} performed poorly with a performance reduction particularly marked in the impact metrics.
The few shots, instead, reduce some of the accuracies without creating a sensitive increase in the others.
Moreover, asking for one component at a time does not add any benefits since it shows low results at the cost of an increased amount of queries.  
These experiments suggest that \acp{llm} already has all the needed information to compute as best as possible the \ac{cvss} vector, and additional data on the prompt are not recommended. 

\subsection{Embeddings}

Similarly, results for the embeddings architecture are shown in Table~\ref{tab:embeddings-results}.
The employed model for classification is XGBoost~\cite{chen2016xgboost}. 
First of all, we can see how results do not change much when changing the embedding model, even though we can see how nomic is performing slightly worse than the other two which returned almost the same average accuracy with some tiny variations in the components. The best model, all-MiniLM, obtained the best accuracy of 0.89 in the \ac{av} element, while \ac{pr} is the worst element with an accuracy of 0.73.

\begin{figure}[htbp]
    \centering
    \includegraphics[width=.95\columnwidth]{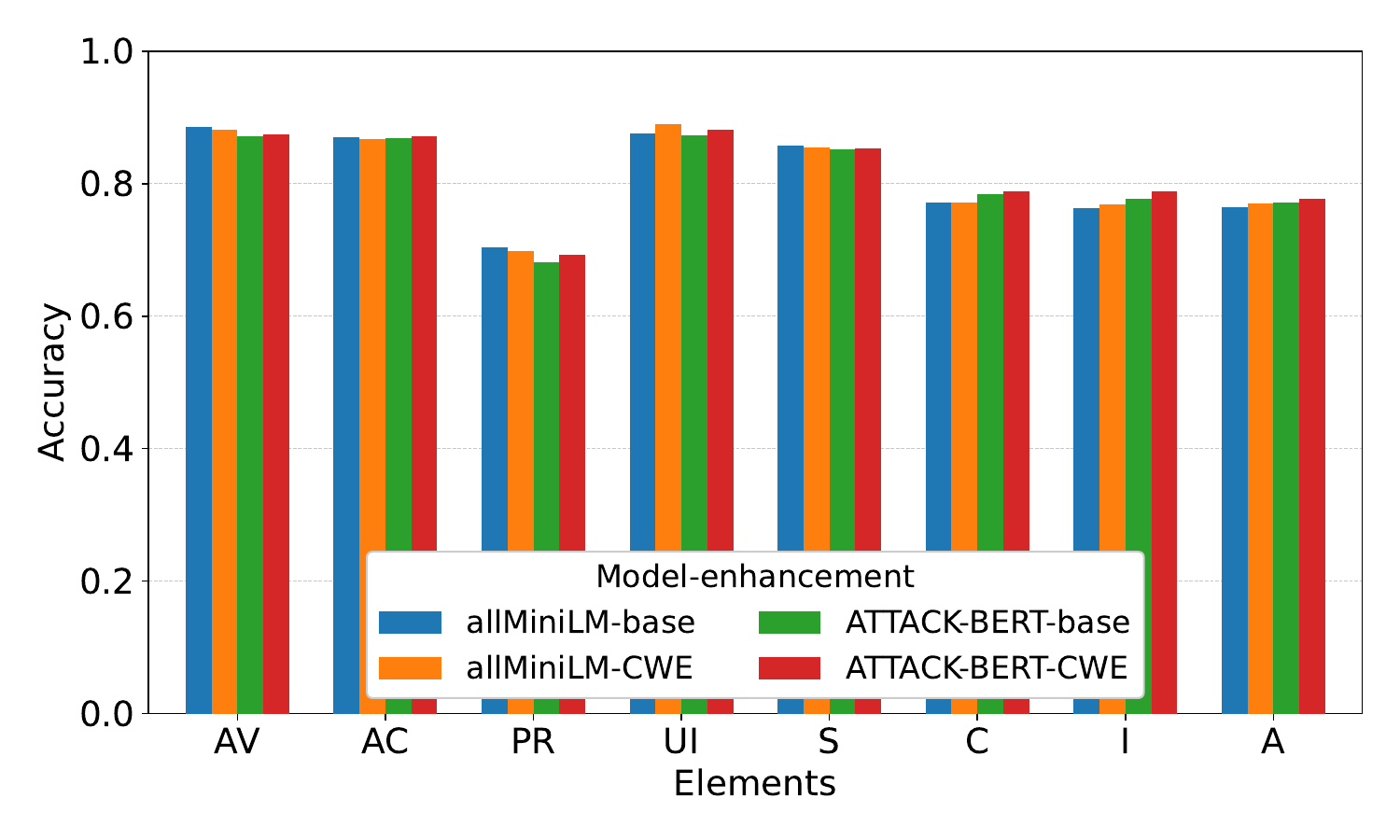}
    \caption{Accuracy provided by base embeddings versus embeddings enhanced with \ac{cwe}.}
    \label{fig:emb-enhancement}
\end{figure}

With respect to the enhancement solutions applied on the prompt in the Vanilla scenario which reduced accuracy, when using embeddings adding \ac{cwe} has only a slight effect on the final accuracy, as shown in Figure~\ref{fig:emb-enhancement}. Due to this negligible improvement, we chose as default the base version which reduced the computation required.

\begin{table}[htbp]
\centering
\caption{Accuracies in the embeddings scenario for different embeddings models in classifying each vector element alone.}
\label{tab:embeddings-results}
\resizebox{\columnwidth}{!}{%
    \begin{tabular}{l|cccccccc} \toprule
    \textbf{Embedding} & \textbf{AV}   & \textbf{AC}   & \textbf{PR}   & \textbf{UI}   & \textbf{S}    & \textbf{C}    & \textbf{I}    & \textbf{A}    \\ \midrule
nomic              & 0.87          & \textbf{0.87} & 0.68          & 0.87          & 0.85          & \textbf{0.78} & \textbf{0.78} & \textbf{0.77} \\
all-MiniLM         & \textbf{0.89} & \textbf{0.87} & \textbf{0.73} & 0.87          & \textbf{0.86} & 0.77          & 0.77          & \textbf{0.77} \\
attack-bert        & \textbf{0.89} & \textbf{0.87} & 0.70          & \textbf{0.88} & \textbf{0.86} & 0.77          & 0.76          & 0.76
    \\ \bottomrule
    \end{tabular}%
}
\end{table}

\subsection{Comparison}
\begin{figure}[htbp]
    \centering
    \includegraphics[width=.95\columnwidth]{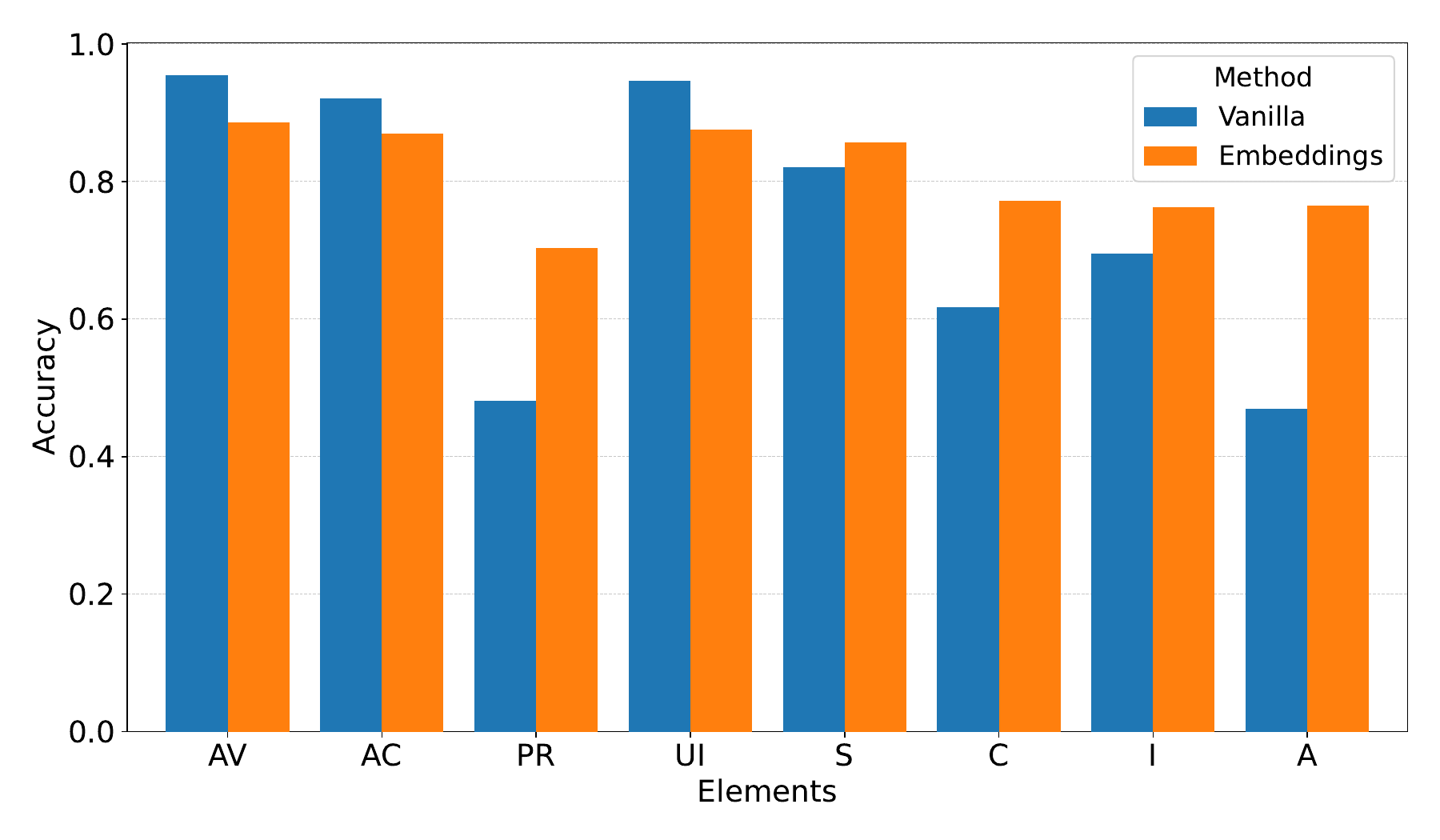}
    \caption{Comparison between the best vanilla (Gemma 3) and the best embedding model (all-MiniLM + XGBoost).}
    \label{fig:comparison}
\end{figure}

As depicted in Section~\ref{sec:methodology}, different enhancements could strengthen the results and have been applied to our scenarios.
Results are shown in Figure~\ref{fig:comparison}.
We show the results for the best \ac{llm} of the vanilla scenario (i.e., Gemma 3) and the best embedding model (i.e., all-MiniLM). 

While results for singular vector elements depend on the element itself, in the vanilla case we can see a higher variation with respect to the embedding scenario. 
While \ac{av} shows lower results than \acp{llm}, other elements where language models struggle to provide a good classification show instead reasonable results in this scenario. An example is, for instance, the \ac{pr} element, where accuracy increases from 0.48 to 0.73. Another interesting result relates to the accuracy for the \ac{a} impact, with an increase from 0.47 to 0.77.
In other cases, such as \ac{av}, \ac{ac}, and \ac{ui}, the accuracy is higher when using \acp{llm}.
Table~\ref{tab:comparison} shows the mean accuracy over all the vector components for the already presented solutions, in addition to a hybrid approach. In this solution, we generate \ac{av}, \ac{ac}, and \ac{ui} from \acp{llm}, while \ac{pr}, \ac{s}, and the impact metrics are generated through embeddings. We observe an improvement in the final accuracy with this method, indicating that a hybrid strategy yields the best results.

\begin{table}[htbp]
\caption{Accuracy in the different proposed approaches.}
\label{tab:comparison}
\centering
\begin{tabular}{lc} \toprule
\textbf{Approach}  & \textbf{Mean Accuracy} \\ \midrule
Vanilla    & 0.738    \\
Embeddings & 0.811    \\
\textbf{Hybrid}     & \textbf{0.835}   
\\ \bottomrule
\end{tabular}
\end{table}

\subsection{Direct Scoring}

As an alternative to component-wise classification, we also experimented with direct \ac{cvss} scoring by prompting language models to estimate the overall score of a vulnerability.
While this method provides less interpretability compared to decomposing the vector, it offers insights into how well \acp{llm} understand and assess vulnerability criticality.
This task serves as a proxy to evaluate their ability to rank \acp{cve} for prioritization. Table~\ref{tab:direct_scoring} reports the Mean Squared Error (MSE) and Mean Absolute Error (MAE) for each model.
Results show that all models deviate from the ground truth, but Qwen 2.5 and GPT 4o yield more accurate scores, indicating stronger potential for scoring tasks.

\begin{table}[htbp]
\caption{Direct CVSS scoring performance of LLMs.}
\label{tab:direct_scoring}
\centering
\begin{tabular}{lcc} \toprule
\textbf{Model} & \textbf{MSE} & \textbf{MAE} \\ \midrule
DeepSeek       & 32.84       & 3.51        \\
Gemma 3        & 2.65        & 1.24        \\
GPT 4o Mini         & 2.47        & 1.21        \\
Llama 3.2      & 3.07        & 1.33        \\
Mistral        & 3.13        & 1.32        \\
Phi 4          & 2.62        & 1.21        \\
Qwen 2.5       & \textbf{2.32} & \textbf{1.06} \\
\bottomrule
\end{tabular}
\end{table}
\section{Discussion}\label{sec:discussion}

The results presented in Section~\ref{sec:results} need some considerations. 
First of all, they suggest that a partial affirmative answer to \textbf{RQ1}, since \acp{llm} can be safely employed to generate vector elements such as \ac{av}, \ac{ac}, and \ac{ui} with a good confidence thanks to accuracies surpassing 0.90. 
On the contrary, \acp{llm} struggle with other elements such as \ac{pr} and the \ac{a} impact, where accuracies below 0.50 do not offer a reliable classification. 
 
We tried different enhancements on the prompt to answer \textbf{RQ2}. As shown in Figure~\ref{fig:prompt-eng}, results were generally not improving with more complex prompts, and often getting worse.
This suggested that a base prompt is the best choice to provide high accuracies and to limit the queries needed to generate each vector. Similarly, even in the embedding scenario adding \ac{cwe} information is not needed to achieve good accuracies. 

The comparison in Figure~\ref{fig:comparison} allows answering \textbf{RQ3}.
However, the answer depends on the considered vector element.
For \ac{av}, \ac{ac}, and \ac{ui} there is a significant increase in the results using \acp{llm}.
However, the other elements do not show the same trend, and the embedding approach results in better results, suggesting that a hybrid approach could be the best solution, as shown in Table~\ref{tab:comparison}. 

Although one might expect \acp{llm} to excel at interpreting subjective aspects due to their ability to reason through context and nuance, this doesn't always translate into accurate classification. Metrics like “Low” vs. “High” confidentiality (\ac{c}) or integrity (\ac{i}) impacts are inherently vague and domain-dependent. Even official \ac{cvss} v3.1 scores can vary between analysts, highlighting the subjective nature of the task~\cite{wunder2024shedding}.

Moreover, \acp{llm} may also overfit to vague interpretations or rhetorical cues not closely tied to \ac{cvss} definitions.
In contrast, embedding-based models, particularly when fine-tuned on security data, tend to capture more subtle patterns linked to severity, suggesting improved alignment with statistical regularities. 
This subjectivity extends to several \ac{cvss} vector elements.
While some, like \ac{av}, are more objective, others (such as \ac{pr}) depend heavily on analyst judgment.
Different researchers may weigh factors differently, leading to inconsistent classifications.

Additionally, human analysts often struggle with accurate \ac{cvss} scoring~\cite{wunder2024shedding}, resulting in mislabeled \acp{cve}. In such cases, \acp{llm} or embedding-based models may even outperform the ``ground truth'', suggesting that generative models could eventually exceed human performance---a possibility worth exploring further.
\section{Conclusions}\label{sec:conclusions}

This study explored the potential of \acp{llm} in automating the generation of \ac{cvss} vectors from \ac{cve} descriptions, a task that is increasingly vital due to the growing volume of reported vulnerabilities. Our experiments demonstrated that while \acp{llm} can achieve high accuracy on certain \ac{cvss} elements, particularly those with more objective criteria, they struggle with more subjective dimensions compared to traditional embedding-based methods. 
Notably, a hybrid approach that combines \acp{llm} with embedding models proved to be a promising direction, offering improved overall performance.
These findings underscore the challenges of \ac{cvss} scoring and suggest refining hybrid models with domain tuning and analyst feedback to improve real-world reliability.
Moreover, in this research, we only considered the most widely used elements of the \ac{cvss} metric. However, temporal and environmental metrics are also available~\cite{cvss-specifications}, even if not mandatory and rarely employed. Especially for this reason, \acp{llm} can represent a good solution to fill their values without augmenting the burden on the user and will be investigated in further studies.

\bibliographystyle{IEEEtran}
\bibliography{bibliography}

\begin{thebibliography}{10}
\providecommand{\url}[1]{#1}
\csname url@samestyle\endcsname
\providecommand{\newblock}{\relax}
\providecommand{\bibinfo}[2]{#2}
\providecommand{\BIBentrySTDinterwordspacing}{\spaceskip=0pt\relax}
\providecommand{\BIBentryALTinterwordstretchfactor}{4}
\providecommand{\BIBentryALTinterwordspacing}{\spaceskip=\fontdimen2\font plus
\BIBentryALTinterwordstretchfactor\fontdimen3\font minus \fontdimen4\font\relax}
\providecommand{\BIBforeignlanguage}[2]{{%
\expandafter\ifx\csname l@#1\endcsname\relax
\typeout{** WARNING: IEEEtran.bst: No hyphenation pattern has been}%
\typeout{** loaded for the language `#1'. Using the pattern for}%
\typeout{** the default language instead.}%
\else
\language=\csname l@#1\endcsname
\fi
#2}}
\providecommand{\BIBdecl}{\relax}
\BIBdecl

\bibitem{wunder2024shedding}
J.~Wunder, A.~Kurtz, C.~Eichenm{\"u}ller, F.~Gassmann, and Z.~Benenson, ``Shedding light on cvss scoring inconsistencies: A user-centric study on evaluating widespread security vulnerabilities,'' in \emph{2024 IEEE Symposium on Security and Privacy (SP)}.\hskip 1em plus 0.5em minus 0.4em\relax IEEE, 2024, pp. 1102--1121.

\bibitem{Kaaviya2025}
\BIBentryALTinterwordspacing
Kaaviya, ``Over 40,000 cves published in 2024, marking a 38\% increase from 2023,'' \emph{Cyber Press}, 2025. [Online]. Available: \url{https://cyberpress.org/over-40000-cves-published-in-2024/}
\BIBentrySTDinterwordspacing

\bibitem{Fortress2024}
\BIBentryALTinterwordspacing
{Fortress Information Security}, ``Nvd analysis report,'' 2024, accessed: 2025-04-08. [Online]. Available: \url{https://www.fortressinfosec.com/nvd-analysis-report}
\BIBentrySTDinterwordspacing

\bibitem{ghosh2025cve}
R.~Ghosh, H.-M. von Stockhausen, M.~Schmitt, G.~M. Vasile, S.~K. Karn, and O.~Farri, ``Cve-llm: Ontology-assisted automatic vulnerability evaluation using large language models,'' \emph{arXiv preprint arXiv:2502.15932}, 2025.

\bibitem{zhang2025llms}
J.~Zhang, H.~Bu, H.~Wen, Y.~Liu, H.~Fei, R.~Xi, L.~Li, Y.~Yang, H.~Zhu, and D.~Meng, ``When llms meet cybersecurity: A systematic literature review,'' \emph{Cybersecurity}, vol.~8, no.~1, pp. 1--41, 2025.

\bibitem{mitra2024localintel}
S.~Mitra, S.~Neupane, T.~Chakraborty, S.~Mittal, A.~Piplai, M.~Gaur, and S.~Rahimi, ``Localintel: Generating organizational threat intelligence from global and local cyber knowledge,'' \emph{arXiv preprint arXiv:2401.10036}, 2024.

\bibitem{perrina2023agir}
F.~Perrina, F.~Marchiori, M.~Conti, and N.~V. Verde, ``Agir: Automating cyber threat intelligence reporting with natural language generation,'' in \emph{2023 IEEE International Conference on Big Data (BigData)}.\hskip 1em plus 0.5em minus 0.4em\relax IEEE, 2023, pp. 3053--3062.

\bibitem{siracusano2023time}
G.~Siracusano, D.~Sanvito, R.~Gonzalez, M.~Srinivasan, S.~Kamatchi, W.~Takahashi, M.~Kawakita, T.~Kakumaru, and R.~Bifulco, ``Time for action: Automated analysis of cyber threat intelligence in the wild,'' \emph{arXiv preprint arXiv:2307.10214}, 2023.

\bibitem{zhang2024attackg}
Y.~Zhang, T.~Du, Y.~Ma, X.~Wang, Y.~Xie, G.~Yang, Y.~Lu, and E.-C. Chang, ``Attackg+: Boosting attack knowledge graph construction with large language models,'' \emph{arXiv preprint arXiv:2405.04753}, 2024.

\bibitem{fayyazi2024advancing}
R.~Fayyazi, R.~Taghdimi, and S.~J. Yang, ``Advancing ttp analysis: harnessing the power of large language models with retrieval augmented generation,'' in \emph{2024 Annual Computer Security Applications Conference Workshops (ACSAC Workshops)}.\hskip 1em plus 0.5em minus 0.4em\relax IEEE, 2024, pp. 255--261.

\bibitem{ullah2024llms}
S.~Ullah, M.~Han, S.~Pujar, H.~Pearce, A.~Coskun, and G.~Stringhini, ``Llms cannot reliably identify and reason about security vulnerabilities (yet?): A comprehensive evaluation, framework, and benchmarks,'' in \emph{2024 IEEE Symposium on Security and Privacy (SP)}.\hskip 1em plus 0.5em minus 0.4em\relax IEEE, 2024, pp. 862--880.

\bibitem{lin2023hw}
Y.-Z. Lin, M.~Mamun, M.~A. Chowdhury, S.~Cai, M.~Zhu, B.~S. Latibari, K.~I. Gubbi, N.~N. Bavarsad, A.~Caputo, A.~Sasan \emph{et~al.}, ``Hw-v2w-map: Hardware vulnerability to weakness mapping framework for root cause analysis with gpt-assisted mitigation suggestion,'' \emph{arXiv preprint arXiv:2312.13530}, 2023.

\bibitem{donadel2024can}
D.~Donadel, F.~Marchiori, L.~Pajola, and M.~Conti, ``Can llms understand computer networks? towards a virtual system administrator,'' in \emph{2024 IEEE 49th Conference on Local Computer Networks (LCN)}.\hskip 1em plus 0.5em minus 0.4em\relax IEEE, 2024, pp. 1--10.

\bibitem{shahid2021cvss}
M.~R. Shahid and H.~Debar, ``Cvss-bert: Explainable natural language processing to determine the severity of a computer security vulnerability from its description,'' in \emph{2021 20th IEEE International Conference on Machine Learning and Applications (ICMLA)}.\hskip 1em plus 0.5em minus 0.4em\relax IEEE, 2021, pp. 1600--1607.

\bibitem{das2021v2w}
S.~S. Das, E.~Serra, M.~Halappanavar, A.~Pothen, and E.~Al-Shaer, ``V2w-bert: A framework for effective hierarchical multiclass classification of software vulnerabilities,'' in \emph{2021 IEEE 8th International Conference on Data Science and Advanced Analytics (DSAA)}.\hskip 1em plus 0.5em minus 0.4em\relax IEEE, 2021, pp. 1--12.

\bibitem{cvss-specifications}
C.~Group \emph{et~al.}, ``Common vulnerability scoring system version 3.1: Specification document,'' in \emph{FIRST-Forum of Incident Response and Security Teams, Cary, USA, Standard}, 2019.

\bibitem{kusner2015word}
M.~Kusner, Y.~Sun, N.~Kolkin, and K.~Weinberger, ``From word embeddings to document distances,'' in \emph{International conference on machine learning}.\hskip 1em plus 0.5em minus 0.4em\relax PMLR, 2015, pp. 957--966.

\bibitem{guo2025deepseek}
D.~Guo, D.~Yang, H.~Zhang, J.~Song, R.~Zhang, R.~Xu, Q.~Zhu, S.~Ma, P.~Wang, X.~Bi \emph{et~al.}, ``Deepseek-r1: Incentivizing reasoning capability in llms via reinforcement learning,'' \emph{arXiv preprint arXiv:2501.12948}, 2025.

\bibitem{team2024gemma}
G.~Team, T.~Mesnard, C.~Hardin, R.~Dadashi, S.~Bhupatiraju, S.~Pathak, L.~Sifre, M.~Rivi{\`e}re, M.~S. Kale, J.~Love \emph{et~al.}, ``Gemma: Open models based on gemini research and technology,'' \emph{arXiv preprint arXiv:2403.08295}, 2024.

\bibitem{hurst2024gpt}
A.~Hurst, A.~Lerer, A.~P. Goucher, A.~Perelman, A.~Ramesh, A.~Clark, A.~Ostrow, A.~Welihinda, A.~Hayes, A.~Radford \emph{et~al.}, ``Gpt-4o system card,'' \emph{arXiv preprint arXiv:2410.21276}, 2024.

\bibitem{touvron2023llama}
H.~Touvron, T.~Lavril, G.~Izacard, X.~Martinet, M.-A. Lachaux, T.~Lacroix, B.~Rozi{\`e}re, N.~Goyal, E.~Hambro, F.~Azhar \emph{et~al.}, ``Llama: Open and efficient foundation language models,'' \emph{arXiv preprint arXiv:2302.13971}, 2023.

\bibitem{jiang2024mixtral}
A.~Q. Jiang, A.~Sablayrolles, A.~Roux, A.~Mensch, B.~Savary, C.~Bamford, D.~S. Chaplot, D.~d.~l. Casas, E.~B. Hanna, F.~Bressand \emph{et~al.}, ``Mixtral of experts,'' \emph{arXiv preprint arXiv:2401.04088}, 2024.

\bibitem{abdin2024phi}
M.~Abdin, J.~Aneja, H.~Behl, S.~Bubeck, R.~Eldan, S.~Gunasekar, M.~Harrison, R.~J. Hewett, M.~Javaheripi, P.~Kauffmann \emph{et~al.}, ``Phi-4 technical report,'' \emph{arXiv preprint arXiv:2412.08905}, 2024.

\bibitem{yang2024qwen2}
A.~Yang, B.~Yang, B.~Zhang, B.~Hui, B.~Zheng, B.~Yu, C.~Li, D.~Liu, F.~Huang, H.~Wei \emph{et~al.}, ``Qwen2. 5 technical report,'' \emph{arXiv preprint arXiv:2412.15115}, 2024.

\bibitem{chen2016xgboost}
T.~Chen and C.~Guestrin, ``Xgboost: A scalable tree boosting system,'' in \emph{Proceedings of the 22nd acm sigkdd international conference on knowledge discovery and data mining}, 2016, pp. 785--794.

\end{thebibliography}

\end{document}